\documentclass[12pt]{article}
\usepackage{amsmath}
\usepackage{multirow}
\usepackage{amsfonts}
\usepackage{amssymb,graphics,psfrag}
\usepackage{array,epsfig,multirow,stmaryrd,graphicx}
\usepackage{comment}

\usepackage{slashed}

\usepackage{hyperref}

\def\hybrid{\topmargin -20pt    \oddsidemargin 0pt
        \headheight 0pt \headsep 0pt 
        \textwidth 6.25in      
        \textheight 9 in      
        \marginparwidth .875in
        \parskip 5pt plus 1pt
          \jot = 1.5ex
  }
\hybrid
\numberwithin{equation}{section}
\numberwithin{table}{section}

\newcommand{\beq}{\begin{equation}}
\newcommand{\eeq}{\end{equation}}
\newcommand{\be}{\begin{equation}}
\newcommand{\ee}{\end{equation}}
\newcommand{\bea}{\begin{eqnarray}}
\newcommand{\eea}{\end{eqnarray}}
\newcommand{\ben}{\begin{eqnarray*}}
\newcommand{\een}{\end{eqnarray*}}               
\newcommand{\ba}{\begin{aligned}}
\newcommand{\ea}{\end{aligned}}
\newcommand{\bt}{\begin{tabular}}
\newcommand{\et}{\end{tabular}}
\newcommand{\bc}{\begin{center}}
\newcommand{\ec}{\end{center}}

\newcommand{\cN}{\mathcal{N}}

\newcommand{\nn}{\nonumber}

\newcommand{\cref}{{\bf [check ref]}}

\newcommand{\tr}{\mathrm{tr}\, }

\psfrag{n1}{$\nu_1$}
\psfrag{n2}{$\nu_2$}
\psfrag{n1'}{$\nu_1'$}
\psfrag{n2'}{$\nu_2'$}
\psfrag{n9}{$\nu_9$}
\psfrag{n10'}{$\nu_{10}'$}
\psfrag{t1}{${\nu}_1$}
\psfrag{t2}{${\nu}_2$}
\psfrag{t9}{${\nu}_9$}
\psfrag{t1'}{${\nu}_1'$}
\psfrag{t2'}{${\nu}_2'$}
\psfrag{t10'}{${\nu}_{10}'$}

\def\blfootnote{\xdef\@thefnmark{}\@footnotetext}
\long\def\symbolfootnote[#1]#2{\begingroup%
\def\thefootnote{\fnsymbol{footnote}}\footnote[#1]{#2}\endgroup}

\begin{document}

\baselineskip=15pt

\begin{titlepage}
\begin{flushright}
\parbox[t]{1.8in}{\begin{flushright} MPP-2013-50\\CERN-PH-TH/2013-044\end{flushright}}
\end{flushright}

\begin{center}

\vspace*{ 1.2cm}

{\large \bf Exploring 6D origins of 5D supergravities\\[.2cm] 
  with Chern-Simons terms}

\vskip 1.2cm

\begin{center}
 {Federico Bonetti\footnote{bonetti@mpp.mpg.de}, Thomas W.~Grimm\footnote{grimm@mpp.mpg.de} and Stefan Hohenegger\footnote{stefan.hohenegger@cern.ch}}
\end{center}
\vskip .2cm
\renewcommand{\thefootnote}{\arabic{footnote}}

{\footnotemark[1]\,\footnotemark[2] Max-Planck-Institut f\"ur Physik, \\
F\"ohringer Ring 6, 80805 Munich, Germany} 

{\footnotemark[3] Department of Physics, CERN - Theory Division, \\
 CH-1211 Geneva 23, Switzerland} 

 \vspace*{1cm}

\end{center}

\vskip 0.2cm
 
\begin{center} {\bf ABSTRACT } \end{center}
We consider five-dimensional supergravity theories with
eight or sixteen supercharges with Abelian vector fields and ungauged
scalars. We address the question under which conditions these theories can be interpreted
as effective low energy descriptions of circle reductions of anomaly
free six-dimensional theories with (1,0) or (2,0) supersymmetry.
We argue that  classical and one-loop
gauge- and gravitational Chern-Simons terms are instrumental for this question.

\hfill {January, 2013}
\end{titlepage}


\section{Introduction}

In this note we address the  question whether it is possible to determine if  a given five-dimensional
supergravity theory can be understood as the effective 
low-energy description of an anomaly-free six-dimensional
supergravity theory on a circle.
Our investigation is motivated by the following considerations.
On general grounds, it is an interesting problem to study the constraints
that  gravity places  on 
low-energy quantum field theories. 
For instance, even-dimensional chiral theories
are subject to the requirement of cancellation
of gravitational anomalies. 
In the spirit of \cite{Vafa:2005ui,Ooguri:2006in,Kumar:2009us,Taylor:2011wt},
one can maybe look for analogue constraints
in odd-dimensional theories by exploring classes of 
models that cannot be seen as a circle reduction
of an anomaly-free even-dimensional theory.
More specifically, 
 the study of five-dimensional quantum field theories 
 with coupling to gravity has recently attracted a lot of attention,
 partly related to the attempt to find an effective world-volume 
 action for multiple M5-branes \cite{Douglas:2010iu,Lambert:2010iw,Bonetti:2012st,Ho:2011ni,Huang:2012tu}. Given the great 
 number of new insights, it would be desirable to classify 
 those theories which are consistent at the quantum level. This is a formidable 
 task and therefore it is advantageous to first try to understand a subset of these 
 theories, namely those  that come from a circle reduction from six dimensions
 (see figure 1). Of course, not all consistent five-dimensional theories arise in such a 
 circle compactification. Well-known examples include Calabi-Yau threefold 
 reductions of M-theory that in general do not admit a six-dimensional lift 
 if the threefold is not elliptically fibered \cite{Cadavid:1995bk,Ferrara:1996wv,Bonetti:2011mw}.
\begin{figure}[h!]
\begin{center}
\includegraphics[width=17cm]{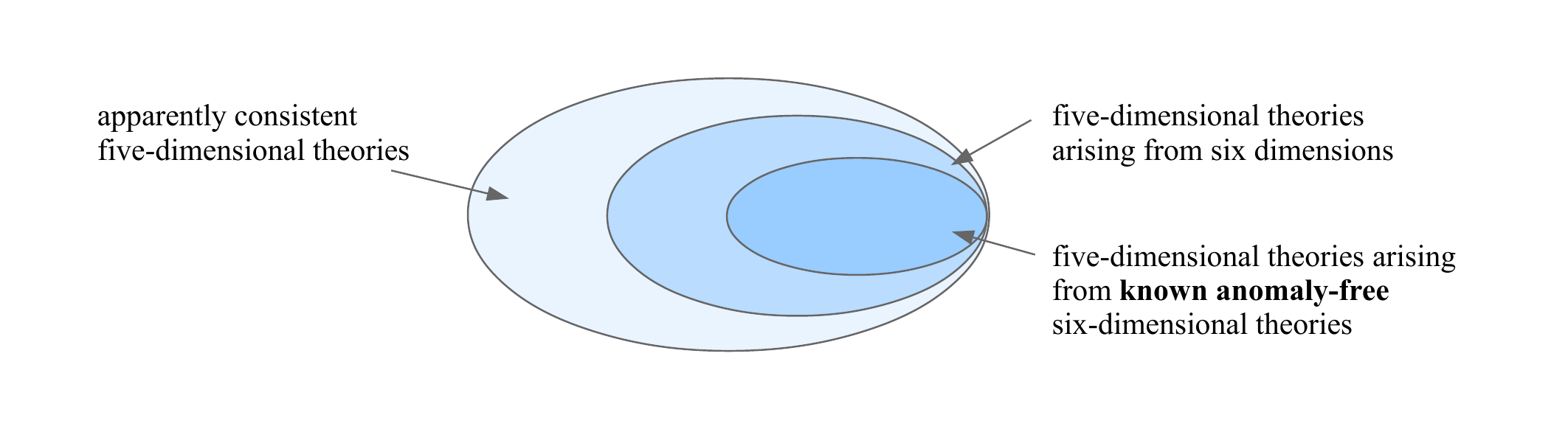}
\vspace{-1cm}
\caption{Five dimensional effective low-energy theories coupled to
gravity which arise through compactification of anomaly-free
six-dimensional theories form a subset of all apparently
quantum-consistent theories.}
\label{Fig:landscape}
\end{center}
\end{figure} 

Deciding upon   this question is generically  a highly non-trivial task,
for various reasons.
On the one hand, in order to extract the low-energy effective action
of a six-dimensional theory on a circle one needs not only 
to perform a classical dimensional reduction, but also
to integrate out massive excitations such as Kaluza-Klein modes.
Five-dimensional quantum effects due to these massive excitations
can   make a direct comparison to a possible higher-dimensional action 
prohibitively difficult.
 On the other hand,
the structure of six-dimensional supergravities
is quite rich and is not completely under control.
The study of non-Abelian interactions among self-dual tensors,
in particular, remains an open problem in the context
of $(2,0)$ theories and has been 
investigated in $(1,0)$ models
in the regime where gravity is decoupled \cite{Samtleben:2011fj}.

Even if we do not have control over the full
class of six-dimensional supergravities,
we can still formulate non-trivial conditions
for a given five-dimensional theory to be lifted
to a specific subset of six-dimensional models.
Moreover, there are objects
at the quantum level of the theory
that are robust under dimensional reduction.
Anomalies, and in particular gravitational ones, 
are examples of such objects, since 
they are mostly sensitive to 
more general features of the theory rather than intricate details of the action 
\cite{AlvarezGaume:1983ig}. 
 In this note, we discuss the possibility to study them using classical and one-loop gauge 
and gravitational Chern-Simons terms in the theory obtained by compactification on a circle. 
Reversing the logic, we try to argue that a careful study of Chern-Simons terms in a generic 
five-dimensional gauge theory allows to obtain non-trivial 
information about the spectrum (and thus also 
about the quantum-consistency) of a potential six-dimensional parent theory. 

The two setups that we investigate admit eight and sixteen supercharges, respectively. 
Firstly, we suppose we are given a five-dimensional
Abelian action with eight supercharges and we
explore the possibility to lift it to a $(1,0)$ theory
with simple gauge group. We find that 
non-trivial necessary conditions 
can be formulated in terms of the Chern-Simons sector only.
Secondly, we take an Abelian theory with sixteen supercharges and we search
for a possible lift to an Abelian $(2,0)$ theory.
As before, a necessary condition on the Chern-Simons couplings,
accompanied by suitable kinetic terms to fix the normalization 
of the fields, is found.

\section{Six-dimensional origin of five-dimensional theories}
\subsection{$\cN=2$ supersymmetric theories}

The minimal amount of supersymmetry in 
five-dimensions consists of eight real supercharges
and will be referred to as $\cN=2$ supersymmetry.
We consider  minimal supergravity   
coupled to $n$ Abelian vector multiplets and a number
of massless neutral hypermultiplets. 
The supersymmetric action 
of such a theory contains the topological 
couplings
\beq \label{gen_CS}
   S^{(5)}_{CS} = 
  \frac{1}{(2\pi)^2}
  \int \bigg[
k_{ABC}\, A^A \wedge F^B \wedge F^C 
    + \,   \kappa_A\, A^A \wedge \tr(R\wedge R)  
  \bigg] \ ,
\eeq
where 
$A^A$, $A = 1, \dots , n +1$ denotes collectively the graviphoton and 
the vectors from the vector multiplets, $F^A = dA^A$ are the corresponding 
 Abelian field strengths, and 
$R$ is the curvature two-form. Supersymmetrizations of the second term are
discussed in \cite{Hanaki:2006pj, Ozkan:2013uk}.

If  an $\cN=2$ theory can be seen as the circle reduction of a six-dimensional theory,
it has to come from a $(1,0)$ theory: For one, if the six-dimensional theory
had more supersymmetry, we would find more than eight supercharges in 
five dimensions.\footnote{Here we consider only simple compactifications 
on a circle. In particular, we do not discuss any compactification 
mechanism which (partially) breaks supersymmetry.}
For another, it seems impossible to lift the five-dimensional gravitino of an $\cN=2$ theory
to a consistent, interacting six-dimensional theory with no supersymmetry.
Note that a five-dimensional theory with massless $U(1)$ gauge fields
can arise as low energy effective action of a possibly non-Abelian
six-dimensional theory on a circle. This is what happens when the gauge group is broken 
to the five-dimensional
Coulomb branch by giving a VEV to the scalars in the five-dimensional
vector multiplets.
 For simplicity,
in the following we study the possibility to lift the five-dimensional
theory to a non-Abelian $(1,0)$ with simple gauge group $G$.
The generalization to semi-simple $G$
is straightforward. The inclusion of $U(1)$ factors is also possible, 
 but would make the analysis of the six-dimensional action and anomalies
more involved.

The first step in the search for a parent six-dimensional theory
is to determine if the five-dimensional spectrum can be lifted to 
six-dimensions.
Five-dimensional hypermultiplets directly lift to six-dimensional hypermultiplets, 
which are allowed in the $(1,0)$ theory. To understand the possible lift 
of the vector sector to six dimensions one has to divide the $n+1$ 
five-dimensional vector fields $A^B$ into three sets:
\begin{itemize}
\item the vector $A^0$ that lifts to the Kaluza-Klein vector in the reduction of the six-dimensional metric on a circle;
\item the vectors $A^\alpha,\, \alpha = 1,\ldots, T+1$ that lift to components of $T$ six-dimensional tensor multiplets and a single tensor in the supergravity multiplet;
\item the vectors $A^i ,\, i=1,\ldots,\text{rank}(G)$ that lift to Cartan elements of six-dimensional gauge  group $G$. 
\end{itemize}
Furthermore, to allow for a consistent six-dimensional parent theory, the constants $k_{ABC}$ and $\kappa_A$ in \eqref{gen_CS} 
have to split in such a way to accommodate the following Chern-Simons 
terms for the above mentioned classes of vector fields
\begin{align} \label{CS_liftable}
  S^{(5)}_{CS }   &= 
  \frac{1}{(2\pi)^2}
  \int \bigg[ - \frac 1 2\Omega_{\alpha \beta} A^0  F^\alpha   F^\beta 
       +\frac 12 b^\alpha \Omega_{\alpha \beta} C_{ij}  \, A^\beta   F^i   F^j
        - \frac 18a^\alpha \Omega_{\alpha \beta} A^\beta   \tr R^2   \bigg]  \\
  &+ \frac{1}{(2\pi)^2} \int \bigg[ 
        k_{0} \, A^0   F^0   F^0 
  + k_{ij} \, A^0   F^i   F^j 
  +  k_{ijk}\,  A^i   F^j   F^k  
  + \kappa_0\,  A^0   \tr R^2 \bigg] \ , \nn
\end{align}
where we suppressed wedge products for brevity.
As discussed for example in \cite{Taylor:2011wt,Bonetti:2011mw},  only
the Chern-Simons terms in the first line  
can be lifted to a classical six-dimensional action,
while the terms in the second line
 cannot be obtained
by classical reduction on a circle.
As it has been shown in \cite{Bonetti:2013ela}, however, such terms can arise at the one-loop level
 by integrating out massive spin-1/2, spin 3/2, or two-forms charged under $A^0$ or $A^i$. 
It is precisely the interplay between these 
two subsets of Chern-Simons terms that allows us to formulate 
necessary conditions for the five-dimensional theory
to come from an anomaly-free  $(1,0)$ theory.

Let us  recall briefly the six-dimensional interpretation of the first
line of terms in \eqref{CS_liftable}.
The constant symmetric matrix $\Omega_{\alpha\beta}$
has signature $(1,T)$ and is identified 
with the $SO(1,T)$ invariant metric associated to the moduli space $SO(1,T)/SO(T)$
of the scalars in the tensor multiplets in six-dimensions.
The matrix $C_{ij}$ is identified with the Cartan matrix
of the gauge group $G$. 
The constant vectors $b^\alpha$ and  $a^\alpha$
contain crucial information about the anomaly
of the six-dimensional parent theory. Indeed, they are the coefficient 
of the Green-Schwarz terms that cancel factorisable anomalies.
Note also that the vector $b^\alpha$ determines
the kinetic term of six-dimensional vectors.

As mentioned above, the requirement of anomaly cancellation in the 
parent $(1,0)$ theory allows us to formulate necessary conditions
on the Chern-Simons terms for the lift to six-dimensions to be possible.
In the following, we focus on six-dimensional gravitational
anomalies, since they do not depend on many details of the charged
hypermultiplet spectrum in six dimensions.
Recall that gravitational anomalies are canceled provided that
\cite{Taylor:2011wt}
\beq \label{6Dgrav_anomalies}
H - V = 273 - 29 \, T \ , \qquad
a^\alpha \Omega_{\alpha \beta} a^\beta = 9 - T \ ,
\eeq
where $T$, $V$, $H$ are the number of six-dimensional
tensor multiplets, vector multiplets, and hypermultiplets, respectively.
To check the first condition in \eqref{6Dgrav_anomalies} directly
we would need to know 
the number of hypermultiplets $H$ in six dimensions. 
This number, however, is in general
different from the number 
of neutral massless hypermultiplets
in five dimensions, since some charged hypermultiplets
become massive after breaking of the gauge group,
and therefore do not appear in the five-dimensional effective action.

This problem can be circumvented by studying the Chern-Simons terms in \eqref{CS_liftable}.
In particular, the couplings $k_{0}$ and $\kappa_0$
encode information about the gravitational anomaly 
cancellation conditions \eqref{6Dgrav_anomalies}.
To see this, recall that 
each massless field in the six-dimensional theory
gives rise to a Kaluza-Klein tower of massive modes in five dimensions.
They are all minimally coupled to  the Kaluza-Klein vector $A^0$,
with charge proportional  to the Kaluza-Klein level.
The massive modes of six-dimensional chiral fermions
and (anti)self-dual tensors  are capable of 
generating the couplings $k_{0}$, $\kappa_0$ in \eqref{CS_liftable}
by running in five-dimensional one-loop diagrams.
The total value of $k_0$, $\kappa_0$
is obtained by summing the contribution of all Kaluza-Klein modes
of all relevant fields.\footnote{The sum over Kaluza-Klein
levels is regularized using the Riemann zeta function.
For instance $\sum_n n \rightarrow \zeta(-1) = - 1/12$.} This sum yields \cite{Bonetti:2013ela}
\beq  \label{k0andkappa0}
k_{0} = \frac{1}{24}(T - 9) \ , \qquad
\kappa_0 = \frac{1}{24}(12-T) \ .
\eeq
These expressions hold under the assumption
that the first condition in \eqref{6Dgrav_anomalies}
is satisfied, but they only involve the number $T$
of tensor multiplets of the theory, which 
can be read off from range of 
the $\alpha$ indices in \eqref{CS_liftable}.
Combining \eqref{k0andkappa0} with the second
condition in \eqref{6Dgrav_anomalies} we  
get the following
   necessary conditions 
   for the Chern-Simons
terms \eqref{CS_liftable} to be lifted to six-dimensional
theory free of gravitational anomalies:
\beq \label{conditions10}
 24\, k_0 =  - a^\alpha \Omega_{\alpha \beta} a^\beta =T-9\ , 
\qquad
24\, \kappa_0  = a^\alpha \Omega_{\alpha \beta} a^\beta + 3 = 12-T\ .
\eeq
These equations encode three
independent requirements and cannot be 
trivially satisfied by rescaling $A^0$ and $A^\alpha$.

One can formulate similar tests on the Chern-Simons
coefficients in \eqref{CS_liftable} to 
check if the candidate parent theory is 
free of purely gauge anomalies.
Such conditions involve a comparison between   
$b^\alpha \Omega_{\alpha \beta} b^\beta$
and the coupling $k_{ijk}$,
which contains crucial information 
 about the six-dimensional charged hypermultiplet
spectrum \cite{Intriligator:1997pq, Grimm:2011fx}. While it was only shown for specific 
examples \cite{Grimm:2011fx}, and not yet in general, that the knowledge of the 
Chern-Simons coefficients allows to check 
cancellation of six-dimensional gauge anomalies, we believe that 
such a statement should hold in general. 
In a similar way, we suspect that conditions involving $a^\alpha \Omega_{\alpha \beta} b^\beta$
and the Chern-Simons coupling $k_{ij}$
can be used to test if the six-dimensional theory
is free of mixed gauge-gravitational anomalies.

\subsection{$\cN=4$ supersymmetric theories}
We can  apply the strategy outlined so far 
also to five-dimensional theories with sixteen supercharges,
denoted $\cN=4$. 
We restrict to the  theory
of $n$ Abelian vector multiplets
coupled to supergravity.
Recall that the $\cN=4$ supergravity multiplet 
contains six vectors. 
Five of them form the $\mathbf{5}$ representation of the $SO(5)_R$ R-symmetry group,
while the sixth one is a singlet.\footnote{This structure is fixed by identifying the five-dimensional gravity multiplet.}
 The singlet will be denoted  
$A^0$, and the remaining ones together with the $n$
gauge fields from the vector multiplets are denoted $A^A,\, A=1,\ldots,n+5$. 
The collective index $A$ is a fundamental $SO(5,n)$ index.
The associated constant metric is denoted $\eta_{AB}$. 
With this notation the topological sector of the
action reads
\beq \label{CS_16susy}
   S^{(5)}_{CS} = 
  \frac{1}{(2\pi)^2}
  \int \bigg[
    - \frac 12 \eta_{AB} \, A^0 \wedge F^B \wedge F^C 
    +   \kappa_0\, A^0 \wedge \tr(R\wedge R) 
  \bigg] \ .
\eeq
To the best of our knowledge it has not been shown that
the gravitational Chern-Simons coupling can be 
supersymmetrized. We will see, however, that in some
circumstances it can be generated at the quantum level
from a six-dimensional theory with sixteen supercharges on a circle.
We thus expect it to be an admissible coupling in the
five-dimensional $\cN =4$ action.

In contrast to the $\cN = 2$ case, 
the Chern-Simons sector of an $\cN=4$
theory is too simple to provide any test 
that cannot be trivially satisfied by means
of rescaling of $A^0$, $A^A$. Therefore,
we also need to record some 
kinetic terms in order to fix this
ambiguity. This requires some additional notation.
Each vector multiplet contributes
five scalars to the spectrum. These $5n$
scalars parametrize the coset space 
$SO(5,n)/SO(5)\times SO(n)$.
This is conveniently described in terms of 
 matrices ${L_A}^i$, ${L_A}^I$,
where $i$, $I$ are fundamental indices 
of $SO(5)$, $SO(n)$ respectively.
These matrices satisfy
\beq
\eta_{AB} = \delta_{ij} {L_A}^i {L_B}^j
 - \delta_{IJ} {L_A}^I {L_B}^J \ , 
 \quad
 G_{AB} =  \delta_{ij} {L_A}^i {L_B}^j
 + \delta_{IJ} {L_A}^I {L_B}^J \ , 
\eeq
where $G_{AB}$ is a non-constant, positive-definite
matrix that enters the gauge coupling function. 
The needed kinetic terms 
are
\beq \label{kinetic_16susy}
S^{(5)}_{\rm kin} = \frac{1}{(2\pi)^2}
\int \bigg[
R*1 - \frac 12 d\sigma \wedge * d\sigma
-\frac 12 e^{2\sigma / \sqrt 6} G_{AB} F^A \wedge * F^B
- \frac 12 e^{-4\sigma / \sqrt 6}  F^0 \wedge * F^0
\bigg] \ ,
\eeq
in which $\sigma$ is the scalar in the gravity multiplet.
The sum $S^{(5)}_{CS} + S^{(5)}_{\rm kin}$ can be 
supersymmetrized since it coincides
with part of the standard form of the five-dimensional $\cN=4$
action as found 
e.g.~in \cite{Dall'Agata:2001vb}, up to field redefinitions.\footnote{More
precisely, we have performed 
an overall  rescaling
of the action, together with the
redefinitions $\sigma_{\rm there} = \sigma_{\rm here} / \sqrt 2$, 
$A^0_{\rm there} = A^0_{\rm here}  / \sqrt 2$,
$A^A_{\rm there} = A^A_{\rm here} / \sqrt 2$. 
Our form of the action is best suited for comparison
between tree-level and one-loop terms.
It is such that the action and the vectors
both have period $2\pi$. It has been 
inferred by deriving $S^{(5)}_{CS}$ from
M-theory on $K3\times T^2$  making use of the
effective action discussed in \cite{Witten:1996md}.}

The five-dimensional $\cN = 4$ theory
under examination
can come from circle reduction 
of a $(2,0)$ or $(1,1)$ theory.
Since $(1,1)$ theories are non-chiral,
we cannot use anomalies as a check
of the quantum consistency of the candidate parent theory.
For this reason, in the rest of this section
we formulate necessary conditions
for the lift of the five-dimensional theory
to a $(2,0)$ theory, and we do not give conditions
for the lift to a $(1,1)$ theory.
Furthermore, since a six-dimensional action for 
non-Abelian $(2,0)$ is not known, we
explore the possibility to lift 
the five-dimensional theory to an Abelian $(2,0)$ theory.

Recall that such a theory has only   tensors
as matter multiplets. Cancellation of gravitational anomalies
requires a number $T = 21$ of them.
This implies that the five-dimensional
theory must have exactly $26$ vectors 
in addition to the singlet $A^0$. This provides a first
elementary check on \eqref{CS_16susy}.
A far less trivial check comes from the 
gravitational Chern-Simons coupling $\kappa_0$.
It cannot be generated by reduction of
the classical Abelian $(2,0)$
action on a circle, and it is rather generated
by one-loop diagrams in which  massive
Kaluza-Klein modes run in the loop.
This coupling has been computed in $\cite{Bonetti:2013ela}$
with the result
\beq \label{20condition}
\kappa_0 = \frac 1 4 \ .
\eeq
If in $S^{(5)}_{CS} + S^{(5)}_{\rm kin}$ a different value of $\kappa_0$ appears,
the theory cannot be lifted to an Abelian
$(2,0)$ theory.

\section{Conclusions}

In this work we explored the space of supersymmetric five-dimensional effective 
theories. We addressed the question whether or not a given theory can 
effectively arise from an anomaly-free six-dimensional theory on a circle at 
low energies. By focusing on a certain class of six-dimensional theories 
we formulated explicit constraints on the spectrum and supersymmetry
content of the six-dimensional theory in terms of the 
five-dimensional Chern-Simons couplings. 
We note that our findings based on Chern-Simons terms alone 
cannot be viewed as a classification of all five-dimensional 
theories that can arise in a circle compactification in 
the spirit of figure \ref{Fig:landscape}. 
However, we provided a setup in which this question can be posed 
systematically and checked for a given example. 
Therefore, we see our work as a first step towards a 
systematic analysis of consistency conditions for five-dimensional quantum field theories in the 
presence of gravity.

It is an interesting task to extend our approach to 
more general six-dimensional theories. In particular, one 
might ask to which extent our results can be used 
to explore the possibility of a lift to a non-Abelian $(2,0)$ theory.
To address this question, a remark is due. 
There are two familiar realization of $(2,0)$ theories
in string theory and M-theory. 
On the one hand, they are the world-volume theory of a stack
of M5 branes.
On the other hand, 
these theories can arise from Type IIB on a singular K3.
In the former setup gravitational anomalies 
on the world-volume of the stack of M5 branes
can be canceled by anomaly inflow from the eleven-dimensional
bulk \cite{Freed:1998tg}. In this way, the number of tensor multiplets
is not restricted to be 21. Note instead that in the Type IIB setup
anomaly inflow is not available, and indeed the theory 
possesses 21 tensor multiplets even in presence of
non-Abelian interactions among the tensors.
We can argue that the condition \eqref{20condition}
is still valid in this case, while it is probably
not required if we allow six-dimensional
gravitational anomalies to be canceled
by inflow from some higher-dimensional bulk theory.
To systematically approach the non-Abelian theory from 
five-dimensions remains an exciting challenge and might 
yield intriguing insights about the nature of $(2,0)$ theories.

One might also hope to apply the same strategy
to theories in other dimensions. In particular, 
the study of circle compactifications from four to three
dimensions can be motivated by the duality of F-theory and M-theory 
compactifications and the match of their effective actions.
In the three-dimensional theory Chern-Simons terms are also 
generated at one loop. It was shown in \cite{Grimm:2011fx,Cvetic:2012xn} that they 
capture information about the four-dimensional chiral spectrum 
and its anomalies. Focusing as in five dimensions on the 
Coulomb branch, the Chern-Simons terms are specified 
by a constant matrix $\Theta_{A B}$ for the 
coupling $\int \Theta_{AB}\, A^A \wedge F^B$. These 
encode both the four-dimensional gaugings of axions, as well 
as the one-loop contributions from integrated out massive matter. 
As in five dimensions this matter includes modes that become massive
in the Coulomb branch and fields that are Kaluza-Klein modes. 
However, in contrast to five dimensions one cannot infer all 
relevant information for the four-dimensional Green-Schwarz 
mechanism from the Chern-Simons terms alone \cite{Grimm:2012yq,Cvetic:2012xn}. 
The four-dimensional analogs of $a^\alpha,b^\alpha$ introduced in 
\eqref{CS_liftable} do not appear in Chern-Simons terms and one needs 
to extend the analysis to other couplings of the effective action.
Including these couplings one could proceed in a similar manner as 
in the five-dimensional case and check if a given three-dimensional 
theory can effectively arise from a four-dimensional 
anomaly-free theory. 

Finally, let us note that it is significantly more complicated
to apply the presented strategy to compactifications 
that are not on circles, but on general higher-dimensional 
geometries. To make any concrete statements about the 
underlying theory one would need finer information about the 
effective actions and their corrections. 
Moreover, when
turning to string theory, also massive extended modes can
arise, correct the effective theory, and induce
dualities.\footnote{Higher-dimensional theories might be distinguished
by the representations of the massive modes \cite{AbouZeid:1999fv}, but 
this does not imply that a distinction can be made on the level of the low-energy effective theories.}
Formulating the
criteria which allow an effective theory to arise from string theory 
is a giant mountain dwarfing the hill climbed in this work.

\subsubsection*{Acknowledgments}

We gratefully acknowledge interesting discussions with Ignatios Antoniadis, Denis Klevers, Dieter L\"ust,
and Wati Taylor. This research was supported by a grant of the Max Planck Society.



\end{document}